\documentclass[fleqn,usenatbib]{mnras}

\usepackage{newtxtext,newtxmath}
\usepackage[T1]{fontenc}
\usepackage{ae,aecompl}

\usepackage{graphicx}	% Including figure files
\usepackage{amsmath}	% Advanced maths commands
\usepackage{amssymb}	% Extra maths symbols

\title[Critical conditions for suppression of $\rm H_2$ cooling]{Direct collapse to supermassive black hole seeds:\\ the critical conditions for suppression of $\rm H_2$ cooling}

\author[Y. Luo et al.]{
Yang Luo,$^{1}$\thanks{E-mail: yangluo@xmu.edu.cn (YL)}
Isaac Shlosman,$^{2,3}$
Kentaro Nagamine$^{3,4,5}$
and Taotao Fang$^{1}$
\\
$^{1}$Department of Astronomy and Institute of Theoretical Physics and Astrophysics, Xiamen University, Xiamen, Fujian 361005, China\\
$^{2}$Department of Physics \& Astronomy, University of Kentucky, Lexington, KY 40506-0055, USA\\
$^{3}$Theoretical Astrophysics, Department of Earth \& Space Science, Osaka University, 1-1 Machikaneyama, Toyonaka, Osaka 560-0043, Japan\\
$^{4}$Department of Physics \& Astronomy, University of Nevada, Las Vegas, NV 89154-4002, USA\\
$^{5}$Kavli-IPMU (WPI), University of Tokyo, 5-1-5 Kashiwanoha, Kashiwa, Chiba, 277-8583, Japan
}

\date{Accepted XXX. Received YYY; in original form ZZZ}

\pubyear{2019}

\begin{document}
\label{firstpage}
\pagerange{\pageref{firstpage}--\pageref{lastpage}}
\maketitle

\begin{abstract}

Observations of high-redshift quasars imply the presence of supermassive black holes (SMBHs) already at $z\sim 7.5$. An appealing and promising pathway to their formation is the direct collapse scenario of a primordial gas in atomic-cooling haloes at $z\sim 10 - 20$, when the $\rm H_{2}$ formation is inhibited by a strong background radiation field, whose intensity exceeds a critical value, $J_{\rm crit}$. To estimate $J_{\rm crit}$, typically, studies have assumed idealized spectra, with a fixed ratio of $\rm H_{2}$ photo-dissociation rate $k_{\rm H_2}$ to the $\rm H^-$ photo-detachment rate $k_{\rm H^-}$. This assumption, however, could be too narrow in scope as the nature of the background radiation field is not known precisely. In this work we argue that the critical condition for suppressing the {\rm H$_2$} cooling in the collapsing gas could be described in a more general way by a combination of $k_{\rm H_2}$ and $k_{\rm H^-}$ parameters, without any additional assumptions about the shape of the underlying radiation spectrum. By performing a series of cosmological zoom-in simulations covering a wide range of relevant $k_{\rm H_2}$ and $k_{\rm H^-}$ parameters, we examine the gas flow by following evolution of basic parameters of the accretion flow. We test under what conditions the gas evolution is dominated by $\rm H_{2}$ and/or atomic cooling. We confirm the existence of a critical curve in the $k_{\rm H_2}-k_{\rm H^-}$ plane, and provide an analytical fit to it. This curve depends on the conditions in the direct collapse, and reveals domains where the atomic cooling dominates over the molecular cooling. Furthermore, we have considered the effect of $\rm H_{2}$ self-shielding on the critical curve, by adopting three methods for the effective column density approximation in $\rm H_{2}$. We find that the estimate of the characteristic lengthscale for shielding can be improved by using $\lambda_{\rm Jeans25}$, which is 0.25 times that of the local Jeans length, which is consistent with previous one-zone modeling.

\end{abstract}


%%%%%%%%%%%%%%%%%%%%%%%%%%%%%%%%%%%%%%%%%%%%%%%%%%

%%%%%%%%%%%%%%%%% BODY OF PAPER %%%%%%%%%%%%%%%%%%

\section{Introduction}

Supermassive black holes (SMBHs) with masses of $\sim 10^9\,\rm M_{\odot}$ have been found in the less than 750\,Myr-old universe, at $z\sim 7.5$, in the midst of quasars \citep{fan03,mortlock11,willott10,wu15,venemans17,banados18}. The origin of these SMBHs is still an open question, and it is not clear how they have managed to grow so quickly after the Big Bang. 

The SMBH seeds can in principle grow via supercritical accretion from stellar mass black holes --- the end products of metal-free Population\,III stars \citep[e.g.,][]{madau14,lupi16,li19}. But this growth rate requires a massive reservoir of accretion matter feeding the Pop\,III remnant over long time intervals, e.g., longer than 100\,Myr \citep[e.g.,][]{begelman06,tanaka09}. Such an option can be realized in the form of a supermassive star \citep{volonteri06,begelman10}, but its existence must be verified in the first place. Another possibility is the runaway collapse of compact stellar clusters, subject to general relativistic effects \citep{Zeldovich65,Shapiro85}, or stellar/gas dynamical evolution of stellar clusters \citep{Begelman78,lupi14}. Their existence at high redshifts is difficult to explain, and their collapse imposes additional conditions on properties of their stellar and gaseous components.

One of the most promising ways to form the SMBH seeds at high redshifts is the direct collapse scenario, which involves the baryonic collapse within dark matter (DM) haloes \citep[e.g.,][]{rees84,haehnelt93,loeb94,bromm03,koushiappas04,begelman06}. In the direct collapse, the virial temperature of DM haloes must exceed the cooling floor of the primordial gas, and the seed black holes as massive as $\sim 10^4 - 10^6\,\rm M_\odot$ can form. The direct collapse paradigm requires that the gas remains atomic to avoid fragmentation, and the high inflow rate is maintained. Furthermore, the infalling gas has to overcome the angular momentum barrier \citep{begelman09}. 

Numerical modeling of an optically-thin phase of a direct collapse has been performed and confirmed previous theoretical estimates. The collapsing gas must be prevented from forming molecular hydrogen  \citep[e.g.][]{shang10,latif13} to avoid fragmentation and formation of the Pop\,III stars \citep[e.g.][]{haiman00,bromm03,wise08,regan09,greif11,choi13,choi15,shlosman16,latif13,latif16}.

Suppression of the H$_2$ cooling requires the presence of a strong ultraviolet (UV) background radiation field, which can dissociate H$_2$. In most studies, the minimum value of the UV intensity to suppress the H$_2$ cooling is denoted by $J_{\rm crit}$, in units of $J_{\rm LW,21} = 10^{-21}\rm erg\,s^{-1}\,cm^{-2}\,Hz^{-1}\,sr^{-1}$. The UV background intensity in the early universe is expected to come from the cosmic star formation \citep[e.g.,][]{greif07,haardt12}. Subsequent analysis has indicated that the background intensity may not be strong enough to reach the required $J_{\rm crit}$  \citep[e.g.][]{ciardi00,ciardi05,dijkstra08,ahn09,holzbauer12}. 

To assure that the collapsing gas within a DM halo will be able to follow the isothermal track, it was claimed that the direct collapse haloes must be located close to the starforming galaxies \citep[e.g.,][]{agarwal12,dijkstra14}, or follow synchronized collapse within two haloes at a small separation \citep{visbal14}, in order to be subject to a strong UV flux. However, attempts to search for such haloes and obtain their population have encountered extreme difficulties \citep[e.g.,][]{yue13,chon17,habouzit16}. The estimate of the number density of direct collapse haloes at $z\sim 10$ exposed to radiation from a nearby starforming galaxy is very sensitive to $J_{\rm crit}$ \citep[e.g.,][]{dijkstra08,agarwal12,dijkstra14,yue14,inayoshi15,yue17,maio19}. A variation by an order of magnitude in $J_{\rm crit}$ can lead to the five orders of magnitude variation in this probability. This emphasizes the need to obtain a more stringent constraint on the value of $J_{\rm crit}$ and on the uncertainties in its determination.

The value of $J_{\rm crit}$ is strongly dependent on the background radiation spectral shape. To suppress the H$_2$ cooling, the H$_2$ fraction could be reduced by the $\rm H^{-}$ photo-detachment, as an alternative mechanism to the $\rm H_{2}$ photo-dissociation. Hence, the value of $J_{\rm crit}$ depends on the relative contribution from $\rm H^{-}$ photo-detachment rate, $k_{\rm H^-}$, and from $\rm H_{2}$ photo-dissociation rate, $k_{\rm H_2}$ \citep[e.g.,][]{sugimura14}. For a given radiation spectral shape, $k_{\rm H^-}$ and $k_{\rm H_2}$ can be calculated from 
\begin{eqnarray}
& k_{\rm H^-} = \alpha \kappa_{\rm H^-}  J_{\rm LW} \\
& k_{\rm H_2} = \beta \kappa_{\rm H_2}   J_{\rm LW},
\end{eqnarray}
where $\kappa_{\rm H^-} \approx 1.1\times 10^{-10}\,\rm s^{-1}$ and $\kappa_{\rm H_2}\approx 1.38\times 10^{-12}\,\rm s^{-1}$ are the rate coefficients of $\rm H^{-}$ photo-detachment and $\rm H_{2}$ photo-dissociation, respectively \citep[e.g.,][]{abel97,glover07b,miyake10}. Dimensionless parameters $\alpha$ and $\beta$ provide the dependence on the radiation spectral shape \citep{glover07b}. When the specific intensity at 13.6\,eV becomes larger than $J_{\rm crit}$, the $\rm H_2$ cooling can be inhibited. 

In most cases, the estimate of $J_{\rm crit}$ is obtained by assuming a particular spectral shape, either a blackbody or a power law. To be representative of dominant Pop\,III stellar populations in galaxies, the radiation field has been modeled as a blackbody with $T_{\star} = 10^5$\,K, hereafter referred to as T5 \citep{omukai01,shang10,hartwig15,inayoshi15}. On the other hand, for stellar populations dominated by the Pop\,II stars, the blackbody with $T_{\star} = 10^4$\,K (i.e., T4) has been used \citep{omukai01,shang10,latif14b,inayoshi15}. More realistic spectral shapes can contain a mixture of various stellar populations, and sources with power-law spectra cannot be ruled out either.

The values of $J_{\rm crit}$, in units of $J_{\rm LW,21}$, obtained in previous studies span a large range, from as low as 20 to as high as $10^5$, depending on the incident radiation spectral shape, and the treatment of the H$_2$ self-shielding \citep[e.g.][]{wolcott-green11,sugimura14,hartwig15,agarwal16,wolcott-green17,dunn18}. 

For the T5 and T4 blackbody spectra, the ratio of $k_{\rm H^-}$ to $k_{\rm H_2}$ is fixed. 
However, realistically, the background radiation spectrum is expected to evolve, and hence both $k_{\rm H^-}$ and $k_{\rm H_2}$ will be changing with time. The hardness of radiation spectra of starforming galaxies will evolve as well, and thus its contribution to the $\rm H_{2}$ dissociation rate \citep{leitherer99,schaerer03,inoue11,sugimura14,visbal15,agarwal16}.
Moreover, trapping of Ly$\alpha$ photons emitted in an optically-thick accretion flow during the direct collapse can affect the gas cooling \citep[e.g.,][]{schleicher10, ge17}, and even photo-detach most of $\rm H^{-}$ \citep[e.g.,][]{johnson17}. Under these conditions, the ratio of $k_{\rm H^-}$ to $k_{\rm H_2}$ cannot be calculated simply.  

Additionally, the value of $J_{\rm crit}$ depends on the treatment of $\rm H_2$ self-shielding \citep{wolcott-green11,hartwig15}. With higher $\rm H_2$ density, the gas becomes optically-thick to the UV radiation, and can be self-shielded from the background radiation. Most numerical simulations used the local Jeans length to calculate the column densities for self-shielding, but this assumption can lead to an overestimate of $J_{\rm crit}$ \citep[e.g.,][]{wolcott-green11}. In three-dimensional (3D) simulations, the self-shielding depends on the direction as well, due to a spatial variation of the gas density and its temperature. This directional dependence can cause a substantial difference in the estimate of $J_{\rm crit}$, between 3D and one-zone simulations \citep[e.g.,][]{shang10}. Additional effects, such as shock capturing and hydrodynamical effects, can become important in 3D simulations, as have been suggested \citep[e.g.,][]{latif14b}. Moreover, the X-rays could increase the hydrogen ionization fraction and the free electron fraction, which promotes the $\rm H^-$ and $\rm H_2$ formation via the electron-catalysed reactions (see Equations \ref{equ:h-} and \ref{equ:h2}). The effect of extragalactic X-ray background could increase the value of $J_{\rm crit}$ by a factor of 3 to 10, depending on the spectral shape of the background UV radiation \citep{inayoshi14,latif15,glover16}. Other uncertainties, like chemical reactions \citep{glover15a}, the rate coefficient of the collisional ionization of hydrogen \citep{glover15}, and anisotropy in the external radiation \citep{regan16} could also introduce an uncertainty of up to a factor of 5 into the determination of $J_{\rm crit}$. 

The critical intensity lacks a unique value, and can be determined by a combination of $k_{\rm H^-}$ and $k_{\rm H_2}$, on a two-dimensional plane. For a given $k_{\rm H^-}$, a critical value of $k_{\rm H_2}$ is expected to exist, above which the $\rm H_{2}$ cooling will be suppressed. A critical curve has been found in the $k_{\rm H^-}$ and $k_{\rm H_2}$ plane for one-zone simulations \citep{agarwal16,wolcott-green17}. This curve provides a more general way of determining the critical conditions, without any assumptions of the shape of the underlying radiation spectrum. But a question remains about the existence of such a curve in 3D simulations. How does the critical curve depend on the spatial variations in the density and temperature in 3D? How does the critical curve from 3D hydrodynamic simulations compares to those obtained from one-zone modeling? What is the effect of the $\rm H_2$ self-shielding approximation on the shape of this curve? In this work, we investigate the dependence of $J_{\rm crit}$ in the 3D cosmological zoom-in simulations on the H$_2$ self-shielding modeling and incident spectral shape, as given by $k_{\rm H^-}$ and $k_{\rm H_2}$ parameters.

This paper is structured as follows. Section\,2 describes the numerical methods used here, the initial cosmological conditions, and chemical network for the molecular gas. Our results are presented in Section\,3. Finally, we discuss and summarize this work in Section\,4.

%%%%%%%%%%%%%%%%%%%%%%%%%%%%%%%%%

\section{Numerical Method}

For simulations of direct collapse within DM haloes, we perform 3D zoom-in cosmological simulations using the Eulerian adaptive mesh refinement (AMR) code Enzo-2.5 \citep{bryan97,norman99,bryan14}. To calculate the gravitational dynamics, a particle-mesh $N$-body method is implemented \citep{colella84,bryan95}. The hydrodynamics equations are solved by the piece-wise parabolic solver which is an improved form of the Godunov method \citep{colella84}. It makes use of the particle mesh technique to solve the DM dynamics and the multi-grid Poisson solver to compute the gravity. For more details of our simulations, we refer the reader to \citet{luo16}, \citet{luo18} and \citet{ardaneh18}.

%%%%%%%%%%%%%%%%%%%%%%%%%%%%%%%%%

\subsection{Initial conditions}

We use fully cosmological initial conditions (ICs) for our models and invoke zoom-in simulations \citep[e.g.][]{choi15,luo16,shlosman16,luo18,ardaneh18}. 

For the initial conditions, we adopt the MUSIC algorithm \citep{hahn11}, which uses a real-space convolution approach in conjunction with an adaptive multigrid Poisson solver to generate highly accurate nested density, particle displacement, and velocity fields suitable for multi-scale zoom-in simulations of structure formation in the universe. First, we generate $1\,h^{-1}$ Mpc comoving with root grid $128^3$ DM-only ICs, initially at $z = 199$, and run it without AMR until $z = 10$. Using the HOP group finder \citep{eisenstein98}, we select three haloes with viral temperature above the cooling floor of the primordial gas. Then, we generate a zoom-in DM halo with $1024^3$ effective resolution in DM and gas, centered on the selected halo position. 

The zoom-in region is set to be large enough to cover the initial positions of all selected halo particles. For the DM particles in the zoom-in region, we use 10,223,616 particles which yield an effective DM resolution of about 99\,M$_{\odot}$. The baryon resolution is set by the size of the grid cells. 

The grid cells are adaptively refined based on the following three criteria: baryon mass, DM mass and Jeans length. A region of the simulation grid is refined by a factor of $2$ in length scale, if the gas or DM densities become greater than $\rho_0 2^{\alpha l}$, where $\rho_0$ is the density above which the refinement occurs, $l$ is the refinement level. We set the ENZO parameter {\it MinimumMassForRefinementExponent}
$\alpha$ to $-0.2$, which reduces the threshold for refinement as higher densities are reached. 

We have imposed the condition of at least 16 cells per Jeans length in our simulations, so that no artificial fragmentation would take place \citep{truelove97}. In all simulations, we set the maximum refinement level to $15$, which is about $0.23\ h^{-1}$ pc comoving. 

We use the Planck\,2015 for cosmology parameters \citep{planck-collaboration16}: $\Omega_m$ = 0.3089, $\Omega_\Lambda$ = 0.6911, $\Omega_b$ = 0.04859 $\sigma_8$ = 0.8159, $n_s$ = 0.9667, and $h$ = 0.6774.

%%%%%%%%%%%%%%%%%%%%%%%%%%%%%%%%%

\subsection{Chemical model}

We use the publicly available package GRACKLE-3.1.1\footnote{https://grackle.readthedocs.org/} \citep{bryan14, smith17} to follow thermal and chemical evolution of the collapsing gas. GRACKLE is an open-source chemistry and radiative cooling/heating library suitable for use in numerical astrophysical simulations. 

The rate equations of nine chemical species: $\rm H$, $\rm H^{+}$, $\rm He$, $\rm He^{+}$, $\rm He^{++}$, $\rm e^{-}$, $\rm H^{-}$, $\rm H_{2}$, and $\rm H_{2}^{+}$ are solved self-consistently along with the hydrodynamics in cosmological simulations. In our simulation, we are using the \citet{haardt12} background spectrum. However, it is known that this background intensity is too weak to reach $J_{\rm crit}$ at $z>10$  \citep[e.g.,][]{ciardi05,ahn09,holzbauer12}. The additional local radiation from nearby star formation is considered through the values of $k_{\rm H^-}$ and $k_{\rm H_2}$ in the present work. 
The treatment of $\rm H_{2}$ collisional dissociation by $\rm H$ atom collisions is taken from \citet{martin96} and accounts for both the temperature and density dependence of this process. The rate coefficients for the three-body reaction to form $\rm H_{2}$ is adopted from \citet{forrey13}, which produces a flat temperature dependence. 

We have assumed a dust-free primordial gas and calculated the radiative cooling and heating rates, accounting for collisional excitation, collisional ionization, free-free transitions, recombination, and photoionization heating, depending on the ionizing radiation field. At very high densities, once the $\rm H_{2}$ lines become optically-thick, the decrease in the $\rm H_{2}$ cooling rate is accounted for \citep{ripamonti04}. The collision-induced emission cooling of $\rm H_{2}$ at high densities is also included \citep{ripamonti04}.

In the direct collapse scenario, a crucial assumption is the suppression of the $\rm H_{2}$ formation and cooling. Studies in both semi-analytic analysis and three-dimensional simulations show that a sufficiently strong dissociating Lyman-Werner (LW, $11.2-13.6$ eV) flux is required to suppress the $\rm H_{2}$ cooling entirely. The main pathway for the formation of $\rm H_{2}$ in primordial gas is
\begin{eqnarray}
\mathrm{H + e^{-}} &\rightarrow & \mathrm{H^{-} +}  \gamma \label{equ:h-}\\
\mathrm{ H + H^{-}} &\rightarrow &  \mathrm{H_{2} + e^-.} 
\label{equ:h2}
\end{eqnarray}

In the chemical network, $\rm H_{2}$ can be reduced either by photo-dissociation of  $\rm H_{2}$ or photo-detachment of $\rm H^{-}$. Photo-dissociation of the ground state of $\rm H_{2}$ happens mostly through absorption in the LW bands to the electronically and vibrationally excited states, and then dissociate to the continuum of the ground state, which is known as the Solomon process \citep{stecher67}. On the other hand, $\rm H^{-}$ can be photo-detached by photons with energy above $0.76$\,eV. The chemical reactions are shown as
\begin{eqnarray}
\mathrm{H_{2}} + \gamma_{\rm LW} & \rightarrow & \mathrm{H + H}\\
\mathrm{H^{-}} + \gamma_{0.76} & \rightarrow & \mathrm{H + e^{-}},
\end{eqnarray}
where $\gamma_{\rm LW}$ and $\gamma_{0.76}$ represent the photons in the LW bands and the photons with energy above 0.76\,eV, respectively. In our work, we perform simulations for a set of $\rm H_{2}$ photo-dissociation rate $k_{\rm H_2}$ and $\rm H^{-}$ photo-detachment rate $k_{\rm H^-}$, and find the critical conditions for suppression of the $\rm H_2$ formation. 

%%%%%%%%%%%%%%%%%%%%

\subsection{Numerical self-shielding approximations}

In regions where the LW bands become optically-thick, the photo-dissociation rate and $\rm H_{2}$ abundance are much more suppressed. Therefore the cooling rate depends largely on modeling the self-shielding. Usually a self-shielding factor $f_{\rm sh}$, which is a function of the $\rm H_2$ column density, $N_{\rm H_2}$, is adopted to parameterize the $\rm H_2$ photo-dissociation rate. However, it remains difficult to make an accurate estimate of the self-shielding factor and the $\rm H_2$ column densities. In 3D simulations, it is computationally expensive to find the exact self-shielding column density along the different directions. Alternatively, a local method is used, which relies on the estimate of $N_{\rm H_2}$ from the local properties of the gas, such as $N_{\rm H_2} = n_{\rm H_2} \lambda$, where $n_{\rm H_2}$ is the $\rm H_2$ number density, and $\lambda$ is some characteristic length \citep[e.g.][]{schaye01}. 

\begin{figure*}
\center
\includegraphics[width=.98\textwidth,height=.27\textheight,angle=0]{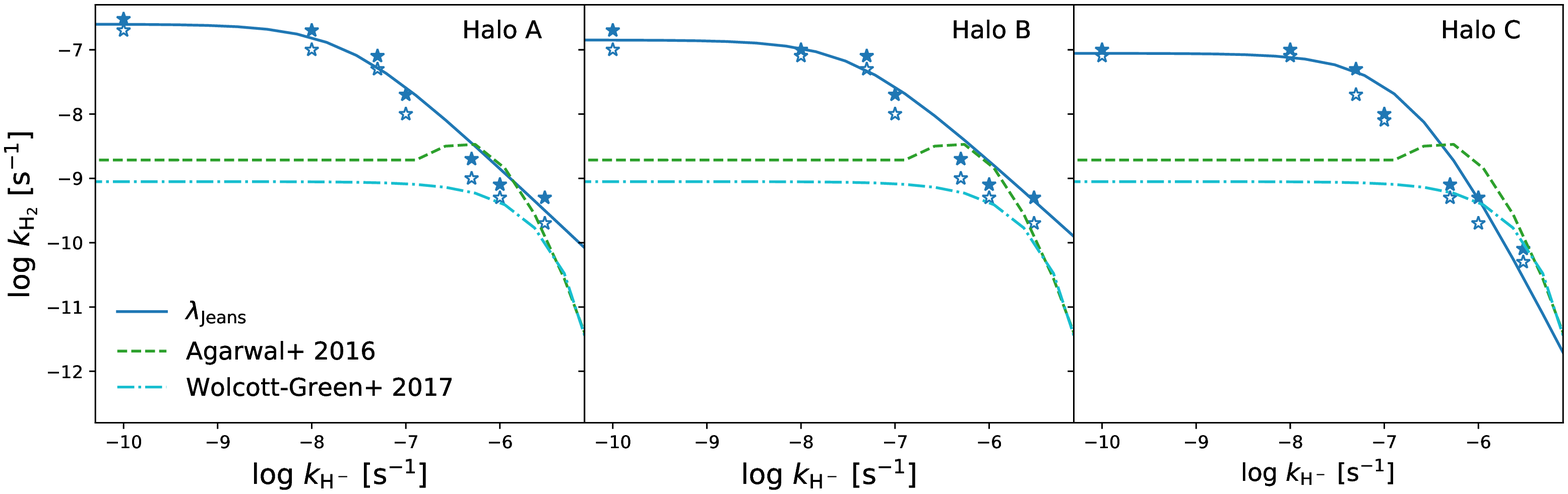}
\caption{Comparison of the critical curve obtained from 3D simulations (blue solid line) with that from one-zone models. One-zone results from \citet{agarwal16} and \citet{wolcott-green17} are shown as green dashed line and cyan dash-dot line, respectively. The green dashed line, as well as our critical blue solid line, are obtained with the same self-shielding approximation $\lambda_{\rm Jeans}$, while the cyan dash-dot line has been obtained with $\lambda_{\rm Jeans25}$. Each line represents the critical boundary in this parameter space above which the $\rm H_2$ cooling is prevented in our models, for haloes A, B and C. Empty and filled marks show the dominant $\rm H_2$ and atomic cooling, respectively, measured in our models. The critical dashed line represents the least square fit to critical points determined by us in the 3D runs.}
\label{fig:crit_curve2}
\end{figure*}

We adopt the improved fitting formula for the self-shielding factor $f_{\rm sh}$ from \citet{wolcott-green11} in our simulations, which is defined as 
\begin{equation}
f_{\rm sh} = \frac{0.965}{(1+x/b_5)^{1.1}} + \frac{0.035}{(1+x)^{0.5}} \exp \left[ - \frac{(1+x)^{0.5}}{1180} \right],
\label{eq:db96}
\end{equation}
where $x=N_{\rm H_2} / 5 \times 10^{14} \rm{cm}^{-2}, b_5 = b/10^5 \rm cm\,s^{-1}$ and $b$ is the Doppler parameter. For the $N_{\rm H_2}$ calculation, the improved approximations, e.g., TreeCol algorithm \citep{clark12,hartwig15a,hartwig15}, six-ray approximation \citep{yoshida07,glover07}, or explicit calculation of the column density using HEALPix \citep{gorski05,regan16}
have been introduced. However, these calculations remain sophisticated and computationally expensive.  Instead, we consider three approximations for the characteristic length  $\lambda$, based on the Jeans length $\lambda_{\rm Jeans}$, Sobolev-like length $\lambda_{\rm Sob}$, and the reduced Jeans length $\lambda_{\rm Jeans25} = 0.25\,\lambda_{\rm Jeans}$ proposed by \citet{wolcott-green17}. The local Jeans length is defined as  $\lambda_{\rm Jeans} = c_{\rm s}\sqrt{\pi/(G\rho)}$, where $c_{\rm s}$ is the sound speed, $G$ is the gravitational constant, and $\rho$ is the gas density. Here $\lambda_{\rm Sob} = \rho/|\nabla \rho|$ depends on the gas density $\rho$ and its spatial gradient. $\lambda_{\rm Sob}$ is a method akin to the Sobolev length and is shown in post-processing of 3D simulations to be accurate in the region where $n_{\rm H_2} < 10^4\ \rm cm^{-3}$ \citep{gnedin09, wolcott-green11}.

%%%%%%%%%%%%%%%%%%%%%%%%%%%%%%%%%%%%%%

\section{Results and Discussion}

\subsection{The critical curve in the 3D simulations}
 
We have generated the ICs for three chosen haloes. For each halo, we run models applying three different self-shielding approximations, and for each approximation, we have calculated a grid of models in the $k_{\rm H_2} - k_{\rm H^-}$ plane. We checked each model for the dominant gas cooling, atomic or $\rm H_2$ as they evolved. 

To examine the effect of $\rm H_2$ formation, we monitor the gas dynamics of the simulated haloes and check evolution of their thermodynamic parameters, including temperature and density. Models dominated by atomic or $\rm H_{2}$ cooling, have a diverging evolution and can be easily distinguished. Fixing the self-shielding approximations and for each $k_{\rm H^-}$, we found pairs of neighboring models with a dominant atomic and H$_2$
cooling along the $k_{\rm H_2}$ axis. The transition or critical point lies between these pairs of models. For each $k_{\rm H^-}$, we have determined the critical points. Thus, we obtained a sequence of critical points which separate the models with atomic and molecular cooling, and which form a critical curve in the $k_{\rm H^-} - k_{\rm H_2}$ plane.

\begin{table}
	\centering
	\caption{Collapse redshift $z_{\rm c}$, dominant cooling $\Lambda$, DM virial mass M$_{\rm v}$, central gas temperature T$_{\rm c}$ and the halo cosmological spin parameter $\lambda_{\rm spin}$ for the simulated haloes, when the maximum refinement level is reached. Here we only list these values for $k_{\rm H^-} = 10^{-8}\,\rm s^{-1}$. The dominant cooling $\Lambda$ refers to models slightly above and below the critical curve shown in Figure\ref{fig:crit_curve2} as filled ({\sc Hi}) and empty (H$_2$) symbols, respectively.}
	\label{table:haloinfo}
	\begin{tabular}{lllllll}
	\hline
	\hline
	  & $\lambda$ & $\Lambda$  & $z_{\rm c}$ & M$_{\rm v}$ & T$_{\rm c}$ & $\lambda_{\rm spin}$\\
	\hline
	Halo A & $\lambda_{\rm Jeans}$   & H$_2$  & 17.1 & 2.5e7 & 798  & 0.03 \\
      		 & $\lambda_{\rm Jeans}$ & {\sc Hi}     & 16.8 & 2.7e7 & 6217 & 0.03 \\
		 
	       & $\lambda_{\rm Jeans25}$ & H$_2$  & 17.3 & 2.3e7 & 871  & 0.03 \\
       		& $\lambda_{\rm Jeans25}$& {\sc Hi}     & 16.8 & 2.7e7 & 6159 & 0.03 \\

	       & $\lambda_{\rm Sob}$     & H$_2$  & 17.3 & 2.3e7 & 389  & 0.03 \\
       		& $\lambda_{\rm Sob}$    & {\sc Hi}     & 16.8 & 2.7e7 & 6142 & 0.03 \\
	\hline	
	Halo B & $\lambda_{\rm Jeans}$   & H$_2$  & 16.7 & 1.7e7 & 839  & 0.01 \\
       		& $\lambda_{\rm Jeans}$  & {\sc Hi}     & 15.8 & 2.20e7 & 6099 & 0.02 \\
			 
	       & $\lambda_{\rm Jeans25}$ & H$_2$  & 18.3 & 1.2e7 & 901  & 0.01 \\
	       & $\lambda_{\rm Jeans25}$ & {\sc Hi}     & 15.7 & 2.2e7 & 6123 & 0.02 \\

	       & $\lambda_{\rm Sob}$     & H$_2$  & 18.0 & 1.2e7 & 482  & 0.00 \\	
	       & $\lambda_{\rm Sob}$     & {\sc Hi}     & 15.8 & 2.2e7 & 6087 & 0.00 \\
	\hline	
	Halo C & $\lambda_{\rm Jeans}$   & H$_2$  & 15.6 & 2.0e7 & 874  & 0.03 \\
       		& $\lambda_{\rm Jeans}$  & {\sc Hi}     & 16.8 & 2.7e7 & 6217 & 0.03 \\
		 
	       & $\lambda_{\rm Jeans25}$ & H$_2$  & 15.7 & 2.0e7 & 883  & 0.03 \\
	       & $\lambda_{\rm Jeans25}$ & {\sc Hi}     & 14.7 & 2.5e7 & 6143 & 0.03 \\

	       & $\lambda_{\rm Sob}$     & H$_2$  & 15.8 & 1.9e7 & 831  & 0.03 \\
	       & $\lambda_{\rm Sob}$     & {\sc Hi}     & 14.8 & 2.4e7 & 6210 & 0.03 \\
	\hline
	\end{tabular}
	\end{table}

As we have discussed in the Introduction, a critical curve in the $k_{\rm H_2} - k_{\rm H^-}$ plane can provide a better description for the $\rm H_{2}$ suppression \citep{agarwal16,wolcott-green17}. Figure\,\ref{fig:crit_curve2} displays our results in the $k_{\rm H_2} - k_{\rm H^-}$ plane, for three different haloes. The empty and filled star symbols represent models dominated by molecular and atomic cooling, respectively, using the self-shielding approximation $\lambda_{\rm Jeans}$.  

Note that models of the same halo and with the same value of $k_{\rm H^-}$ and the shielding parameter $\lambda$, which differ only with $k_{\rm H_2}$ will collapse at slightly different redshift. In Table\,\ref{table:haloinfo}, we list the collapse redshifts $z_{\rm c}$, DM virial masses $M_{\rm v}$, the central gas temperatures $T_{\rm c}$, and the halo cosmological spin parameter $\lambda_{\rm spin}$, for the selected models, measured when the maximum refinement level is reached. The gas collapse proceeds from inside out and leads to a central runaway, and this runaway occurs after the gas density exceeds the background DM density \citep{choi13,choi15,shlosman16}. The collapse proceeds very rapidly, and the maximum refinement level is reached only in about a few million years \citep{luo16}. We stop the simulations when the maximum refinement level has been reached, and measure the required parameters in Table\,\ref{table:haloinfo}. These are listed for different $\lambda$ approximations and the dominant cooling mechanisms for each approximation. Only models with $k_{\rm H^-} = 10^{-8}\,\rm s^{-1}$ are shown in this Table, for simplicity.

In Table\,\ref{table:haloinfo}, models dominated by atomic cooling, i.e., models with $k_{\rm H_2}$ above the critical value, collapse with a slight delay compared to a corresponding model with molecular cooling below the critical point. For all models, the collapse redshifts range from 17 to 13, and the virial masses are approximately a few times $10^7\,\rm M_\odot$. In the $\rm H_{2}$ cooling models, the central gas temperature drops down to a few$\times 100$\,K, while in the atomic cooling models, the temperature remains roughly constant, around 6,000\,K.

For a given $k_{\rm H^-}$, the $\rm H_2$ formation is gradually inhibited with increasing $k_{\rm H_2}$. After determining the critical points for each $k_{\rm H^-}$, we perform the least-square fit and find that the fitting formula can be approximated by
\begin{equation}
	k_{\rm H_2} = a \left (1+\frac{k_{\rm H^-}}{b}\right)^{\rm c} .
\end{equation}

\begin{figure*}
\center
\includegraphics[width=.95\textwidth,height=.5\textheight,angle=0]{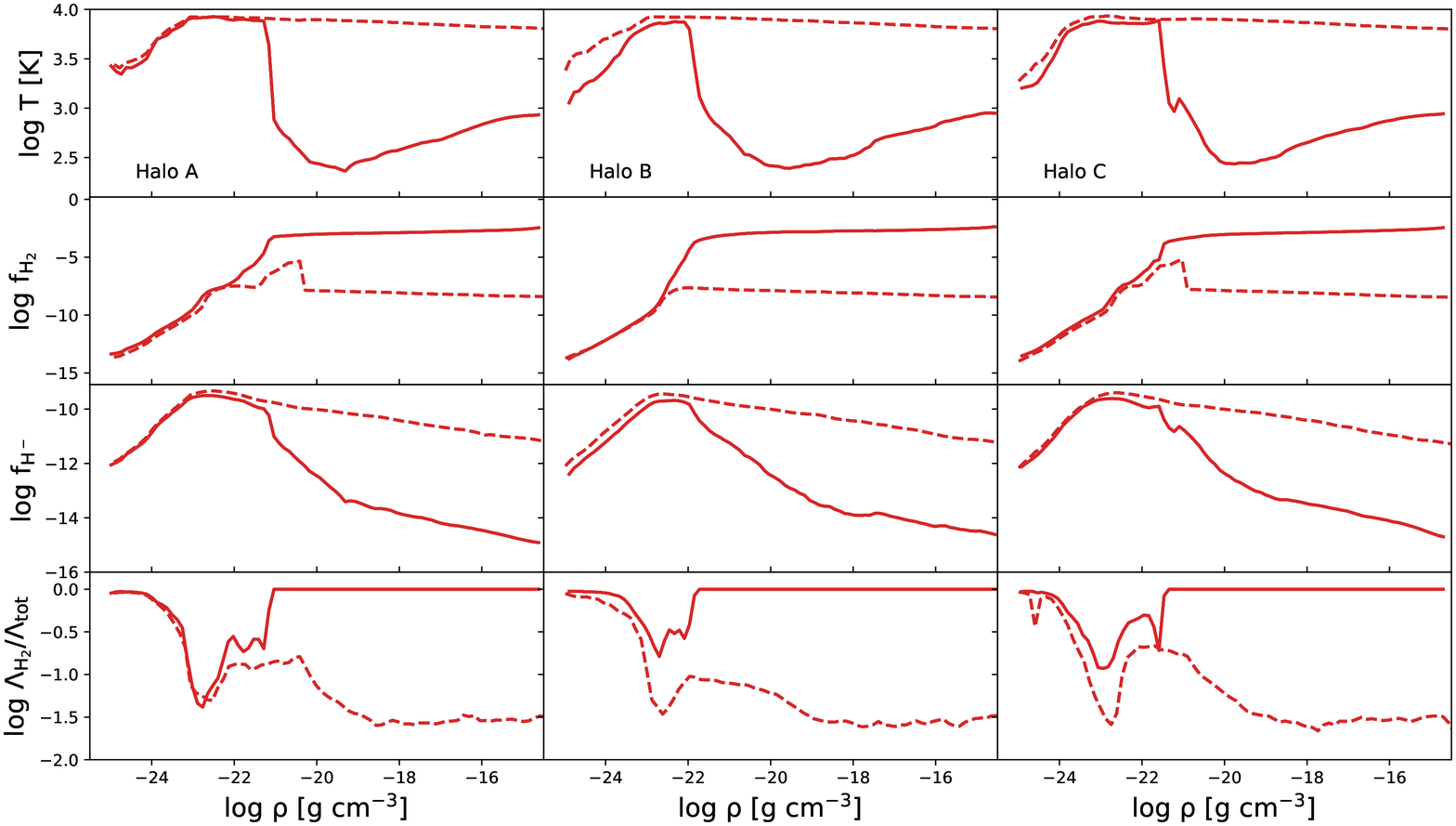}
\caption{Profiles of the gas temperature, of the $\rm H_{2}$ and $\rm H^{-}$ fractions, and the ratios of the molecular to the total cooling rates as a function of the gas density. Only profiles under the condition of $k_{\rm H^-} = 10^{-8}\,\rm s^{-1}$ are shown. The curves represent realizations with $\lambda_{\rm Jeans25}$ approximation, where the solid lines represent the collapse with the dominant $\rm H_{2}$ cooling and the dashed lines represent the collapse with the atomic cooling.}
\label{fig:rhoT}
 \end{figure*}

The fitting parameters have been listed in Table 2. In Figure\,\ref{fig:crit_curve2}, for each halo, we display the fitted critical curve in blue solid line with the self-shielding approximation $\lambda_{\rm Jeans}$. Moreover, we have added the critical curve from \citet{agarwal16} (green dashed line), which has been calculated with the one-zone ENZO code, using the same self-shielding approximation $\lambda_{\rm Jeans}$, and the same cooling package GRACKLE described in Section\,2.2. 

Comparison with the critical curve obtained from the one-zone simulations of \citet{agarwal16} and that from our 3D simulations under otherwise similar conditions, yields a substantial difference between them, up to two orders of magnitude. Furthermore, we have added additional curve (cyan dash-dot line) in Figure\,\ref{fig:crit_curve2} from \citet{wolcott-green17} which has been obtained from one-zone simulations using the cooling package similar to \citet{shang10} and the self-shielding parameter $\lambda_{\rm Jeans25}$. We find the difference between the 3D and one-zone simulations is significant for $k_{\rm H^-} < 10^{-7}\,\rm s^{-1}$. In the vicinity of $k_{\rm H^-} \sim 10^{-6}\,\rm s^{-1}$, the difference minimizes, and then increases again for higher values of $k_{\rm H^-}$. Therefore, the $J_{\rm crit}$ value based on one-zone results could be underestimated (see Section \ref{sect:jcrit} for the $J_{\rm crit}$ calculations). Our results indicate that to suppress the $\rm H_{2}$ cooling, requires a higher LW flux for the same $k_{\rm H^-}$ rate.

Where exactly the evolution of our models with atomic and molecular cooling bifurcates?  Why do models based on the 3D simulations differ profoundly from those in one-zone? We discuss the details of this diverging evolution in the following sections.

\subsection{Impact of the self-shielding column density}

Evolution of direct collapse models depends strongly on the dominant cooling mechanism, which affects their temperature profiles and other thermodynamic parameters. Figure\,\ref{fig:rhoT} exhibits the profiles of the gas temperature, the $\rm H_{2}$ fraction $f_{\rm H_2}$, the $\rm H^{-}$ fraction $f_{\rm H^-}$, and the ratios of the molecular to the total cooling rates, as functions of the gas density, for the self-shielding parameter $\lambda_{\rm Jeans25}$. Additional approximations, $\lambda_{\rm Jeans}$ and $\lambda_{\rm Sob}$, have been evolved as well, but are not shown here. The solid lines show the collapse dominated by the molecular cooling, and the dashed lines correspond to the dominant atomic cooling. 

At the initial stage of the collapse, the gas basically goes into the free-fall, and is shock-heated to the halo virial temperature of $\sim 10^4$\,K around the density of $\rm \sim 10^{-24}\,g\,cm^{-3}$ at the virial radius. When the gas density reaches about $\rm 10^{-23}\,g\,cm^{-3}$, the $\rm H_{2}$ formation becomes important for the future evolution of the gas. However, for higher $k_{\rm H_2}$, the photo-dissociation will suppress the $\rm H_{2}$ formation. Hence, the gas still follows the atomic cooling and the collapse proceeds isothermally. The temperature stays nearly constant around 6,000\,K, and the $\rm H_{2}$ fraction is kept around its the maximum value of $10^{-8}$ only. 

As the gas flows inwards, it remains largely neutral, and the already small $\rm H^{-}$ fraction is slightly decreasing with an increasing density. This is the result of a decreasing fraction of the free electrons required for the $\rm H^{-}$ formation. In cases with the $\rm H_{2}$ cooling being dominant, the total gas cooling rate increases dramatically, causing a substantial drop in the gas temperature, the electron and $\rm H^{-}$ fractions. Around $\rm 10^{-18}\,g\,cm^{-3}$, the collisional dissociation of $\rm H_{2}$ begins to suppress the $\rm H_{2}$ cooling. However, the inflow still does not generate the high density/high temperature regime, the so-called `zone of no return' \citep{inayoshi12, fernandez14}. The compressional heating rate of the collapsing gas decreases substantially due to a lower accretion rate, $\dot {\rm M}$, and the sound speed, $c_{\rm s}$ --- these are related simply by $\dot {\rm M}\sim c_{\rm s}^3/$G. So the heating-cooling balance in the collapsing gas remains at the much lower temperature of the molecular gas.

\begin{figure}
\center
\includegraphics[width=.48\textwidth,height=.35\textheight,angle=0]{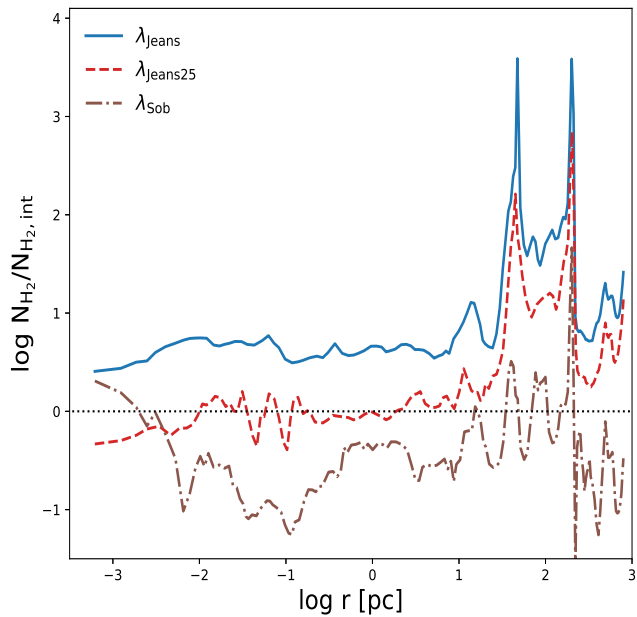}
\caption{Comparison between the column density approximations for halo A. The ratios of the $\rm H_{2}$ column density $N_{\rm H_2}$ to the actual column density $N_{\rm H_2,int}$ are displayed. Approximations for $\lambda_{\rm Jeans}$, $\lambda_{\rm Jeans25}$ and $\lambda_{\rm Sob}$, are shown as the blue solid, red dashed, and brown dash-dot lines, respectively. The black dotted line is drawn to delineate the ratio of unity.}
\label{fig:nh2}
\end{figure}

\begin{figure*}
\center
\includegraphics[width=.9\textwidth,height=.25\textheight,angle=0]{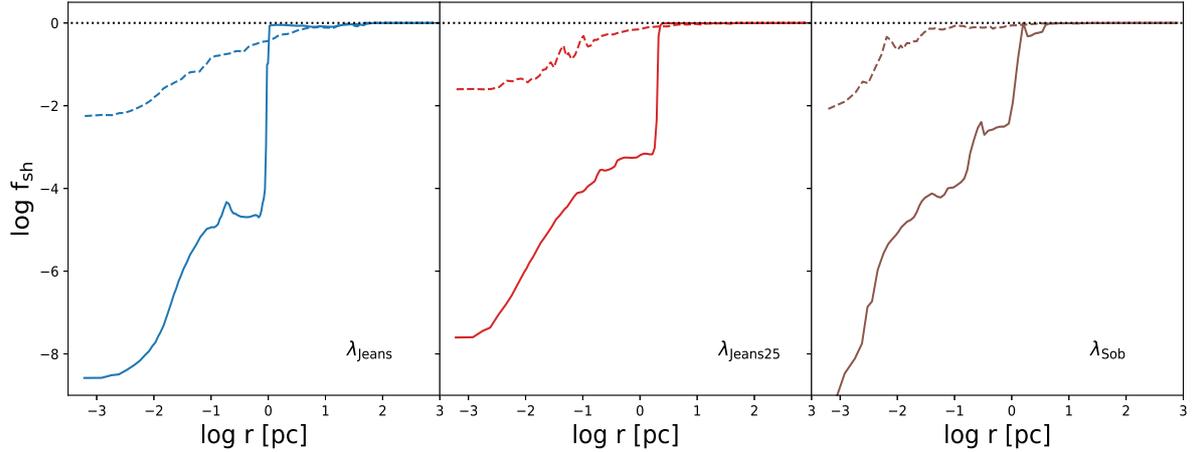}
\caption{Profiles of the self-shielding factor as functions of radius for the simulated halo A at fixed $k_{\rm H^-} = 10^{-8}\,\rm s^{-1}$. The solid lines represent the collapse with the dominant $\rm H_{2}$ cooling and the dashed lines represent the collapse with the dominant atomic cooling. From the left to the right panels, approximations for $\lambda_{\rm Jeans}$, $\lambda_{\rm Jeans25}$ and $\lambda_{\rm Sob}$, are shown as blue, red, and brown lines, respectively. The black dotted line is drawn to delineate the ratio of unity.}
\label{fig:fsh}
\end{figure*}

As we have discussed in the section\,2.3, the primordial gas cooling rate within a DM halo dependents on the self-shielding effect of $\rm H_{2}$. The gas column densities in our models have been calculated using the $\lambda_{\rm Jeans}$, $\lambda_{\rm Jeans25}$ and $\lambda_{\rm Sob}$ approximations. In the next step, we test the accuracy of approximating these column densities, $N_{\rm H_2}$. For comparison, we have calculated the actual column density by integrating the $\rm H_{2}$ profile from the outside inwards, and compared it with column densities obtained from $\lambda_{\rm Jeans}$, $\lambda_{\rm Jeans25}$ and $\lambda_{\rm Sob}$ approximations.  

In Figure\,\ref{fig:nh2}, we present ratios of the column densities $N_{\rm H_2}$ from the adopted self-shielding approximation to that calculated from integrating the $\rm H_{2}$ number density profile. The ratio which stays closer to unity, reflects the more accurate approximation. We display our results for the simulated halo A in the atomic cooling regimes, for fixed $k_{\rm H^-} = 10^{-8}\,\rm s^{-1}$. In this Figure, we only show the results of integration along a single radial sightline in the direction away from the halo center. Results for all three haloes, and for each halo toward different sightlines are qualitatively similar, and so haloes B and C have been omitted from this Figure.
%As shown in Figure\,\ref{fig:nh2}, 
Within the central $\sim$$10$\,pc, the $\lambda_{\rm Jeans}$ approximation gives a higher estimate of $N_{\rm H_2}$ by a factor of 4, while usage of $\lambda_{\rm Sob}$ leads to the estimate which is too low. The $\lambda_{\rm Sob}$ method also shows a relatively large scatter along the radius. However, the ratio obtained from $\lambda_{\rm Jeans25}$ to that determined from simulations lies much closer to unity, within a factor of 2 at radius smaller than 10 pc. We find the $\lambda_{\rm Jeans25}$ approximation provides a more accurate estimate of the $\rm H_{2}$ self-shielding.

The self-shielding factors $f_{\rm sh}$ for halo A along a single sightline are plotted in Figure\,\ref{fig:fsh} for the three approximations, $\lambda_{\rm Jeans}$, $\lambda_{\rm Jeans25}$ and $\lambda_{\rm Sob}$, from left to right, respectively. The solid lines represent the $f_{\rm sh}$ in the $\rm H_2$ cooling cases, and the dashed lines in the atomic cooling cases.
%According to Figure\,\ref{fig:fsh}, 
Outside the central $\sim$$10$\,pc region of the collapse, $f_{\rm sh}$ is insensitive to the choice of $\lambda$. But within the central 10\,pc, the $\rm H_2$ fraction is sharply increasing (see the second row of Figure\,\ref{fig:rhoT}), and the $\lambda$ approximation becomes important for the $f_{\rm sh}$ calculation.

\begin{table}
	\centering
	\caption{Least-square fitting parameters for the critical curves determined from 3D numerical simulations presented in this work (see eq.7).}
	\label{table:fit}
	\begin{tabular}{lllll}
	\hline
	\hline 
	 & $\lambda$ & a  & b & c \\
	\hline
	Halo A & $\lambda_{\rm Jeans}$    & 2.5e-07  & 2.4e-08 & -1.4  \\
       		& $\lambda_{\rm Jeans25}$  & 3.1e-08  & 1.2e-08 & -1.6  \\
			 & $\lambda_{\rm Sob}$      & 1.1e-10  & 1.2e-07 & -1.9  \\
	\hline			 
	Halo B & $\lambda_{\rm Jeans}$    & 1.4e-07  & 3.9e-08 & -1.3  \\
       		& $\lambda_{\rm Jeans25}$  & 3.2e-08  & 1.3e-07 & -1.6  \\
			 & $\lambda_{\rm Sob}$      & 1.1e-10  & 1.1e-07 & -2.0  \\
	\hline			 
	Halo C & $\lambda_{\rm Jeans}$    & 8.8e-08  & 1.9e-07 & -2.9 \\
       		& $\lambda_{\rm Jeans25}$  & 3.2e-08  & 3.1e-07 & -3.1 \\
			 & $\lambda_{\rm Sob}$      & 1.2e-10  & 1.0e-07 & -2.0 \\
	\hline	
	\end{tabular}
\end{table}

\begin{figure*}
\center
\includegraphics[width=.98\textwidth,height=.27\textheight,angle=0]{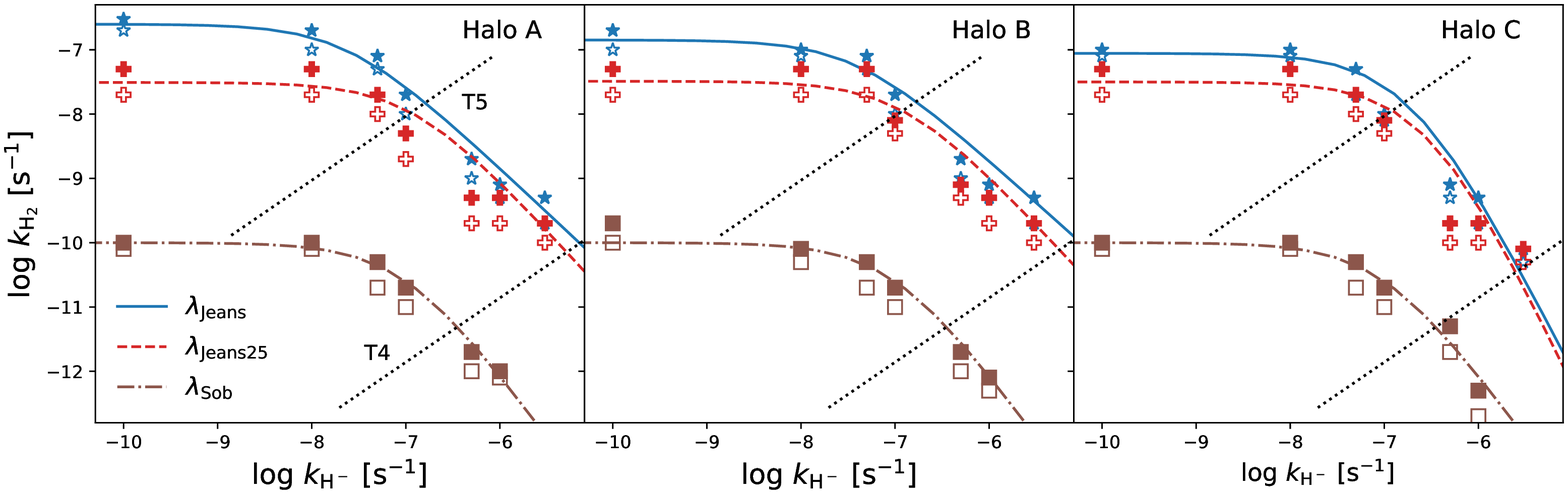}
\caption{The critical curves in the $k_{\rm H_2} - k_{\rm H^-}$ plane (see caption for Figure\,\ref{fig:crit_curve2}), using three different methods to calculate the self-shielding characteristic length $\lambda$; namely, $\lambda_{\rm Jeans}$ (blue solid line, shown also in Fig.\,\ref{fig:crit_curve2}), $\lambda_{\rm Jeans25}$ (red dashed line), and $\lambda_{\rm Sob}$ (brown dash-dot line). The two diagonal black dotted lines represent the rates with varying intensity for the T4 (lower) and T5 spectrum (upper), respectively.}
\label{fig:crit_curve3}
\end{figure*}

Finally, we compared the critical curves calculated in the $k_{\rm H^-} - k_{\rm H_2}$ plane with three different prescriptions for the self-shielding (Figure\,\ref{fig:crit_curve3}). This has been performed for each of the three DM haloes. The fitted curves are given in blue solid, red dashed and brown dash-dot lines for $\lambda_{\rm Jeans}$ (see also Figure\,\ref{fig:crit_curve2}), $\lambda_{\rm Jeans25}$ and $\lambda_{\rm Sob}$, respectively. The least-square fitting parameters have been listed in Table\,\ref{table:fit}.  

The critical curve for $\lambda_{\rm Jeans}$ shows that it lies consistently above that the critical curve for $\lambda_{\rm Sob}$. So the required radiation intensity to suppress the $\rm H_{2}$ formation, therefore, should be stronger as well, which means the required LW radiation intensity is higher for a given $k_{\rm H^-}$. The critical curve is sensitive, by about one order of magnitude, to the choice of the column density approximation used, especially for $k_{\rm H^-} < 10^{-7}\,\rm s^{-1}$. For $k_{\rm H^-} > 10^{-7}\,\rm s^{-1}$, the curve drops down sharply, because  $\rm H^{-}$ photo-detachment becomes the dominant mechanism for the suppression of $\rm H_{2}$ abundance and thus the curve is insensitive to the LW rate $k_{\rm H_2}$. We do not adopt the fitting formula given by the one-zone simulations of \citet{wolcott-green17} and \citet{agarwal16}, which fit the critical curve with an exponential tail at higher $k_{\rm H^-}$. In fact we find that the decay is better fit by a power-law shape, with an index of about $-1.6$. The reason for this is that in the 3D simulations, the spatial variations in the temperature and density within the accretion flow \citep[e.g.,][]{shang10}, or the hydrodynamic effects \citep[e.g.,][]{latif14b} will enhance the $\rm H_{2}$ formation from $\rm H^{-}$. This is a crucial difference between the one-zone and 3D simulations.  

%%%%%%%%%%%%%%%%%%%%%%%%%%%%%%%

\subsection{Calculation of $J_{\rm crit}$ from the critical curve}
\label{sect:jcrit}
In Figure\,\ref{fig:crit_curve3}, the two diagonal dotted lines illustrate the relationship between $k_{\rm H_2}$ and $k_{\rm H^-}$ with varying intensity for a T4 (lower) and T5 (upper) spectral shapes, respectively. We are able to reproduce the critical intensity $J_{\rm crit}$ obtained in previous studies by assuming their fixed blackbody spectral shapes. The critical values lie at the intersection of the diagonal dotted lines crossing the critical curves.

\begin{table*}
	%\begin{center}
	\centering
	\caption{Compilations of $J_{\rm crit}$ values in units of $J_{\rm LW, 21}$ in earlier works using the T5 or T4 black-body spectra. The critical values are obtained at the intersection of the diagonal dotted lines with the critical curves shown in Figures\,\ref{fig:crit_curve2} and \ref{fig:crit_curve3}. The last column lists the method used for the self-shielding approximation.}
	\label{table:jcrit}
	\begin{tabular}{llcccl}
		\hline	
		\hline	
		  Authors & $\lambda$  & J$_{\rm crit, T_5}$ & J$_{\rm crit, T_4}$ & Methods & Approximations\\
		\hline	          
		This Work & $\lambda_{\rm Jeans}$    & 1e4-2e4 & 13-42 & 3D & \citet{wolcott-green11} \\
       			   & $\lambda_{\rm Jeans25}$  & 7e3-8e3 & 13-27 & 3D & \citet{wolcott-green11}\\
			 	   & $\lambda_{\rm Sob}$      & 75-82  & 1.6-1.9  & 3D & \citet{wolcott-green11}\\

		\hline			 
			\citet{shang10}          & $\lambda_{\rm Jeans}$      & 1.2e4    & 39        & one-zone & \citet{draine96}\\
			\citet{shang10}          & $\lambda_{\rm Jeans}$      & 1e4-1e5  & 30-300    & 3D & \citet{draine96}\\
			\citet{latif14b}         &   $--$                     &   $--$   & 400-1500  & 3D & \citet{wolcott-green11}\\
			\citet{latif14b}         &   $--$                     &   $--$   & 30-40     & one-zone & \citet{wolcott-green11}\\
			\citet{sugimura14}       & $\lambda_{\rm Jeans}$      & 1.4e3    & 59.8      & one-zone & \citet{wolcott-green11}\\
			\citet{hartwig15}        & $\lambda_{\rm Jeans}$      & 3.5e3-5.5e3 &    $--$& 3D & \citet{wolcott-green11}\\
			\citet{agarwal16}        & $\lambda_{\rm Jeans}$      & 1.5e3    & 19.4      & one-zone & \citet{wolcott-green11}\\
			\citet{wolcott-green17}  & $\lambda_{\rm Jeans25}$    & 6.6e2    & 19.3      & one-zone & \citet{wolcott-green11}\\

		\hline	
		\end{tabular}
\end{table*}

In Table\,\ref{table:jcrit}, we list the critical intensity $J_{\rm crit}$ for a blackbody spectra of T5 and T4, adopted from previous studies. The second column of Table\,\ref{table:jcrit} displays the self-shielding $\lambda$ approximation adopted in these studies. The third and fourth columns show $J_{\rm crit}$ by assuming a T5 (lower) and T4 (upper) spectral shapes, respectively. The last two columns show the numerical methods used, as well as the methods applied for the self-shielding factor estimates. We have reproduced the calculations of $J_{\rm crit}$. Those lie at the  intersections of the diagonal dotted lines with the critical curves shown in Figures\,\ref{fig:crit_curve2} and \ref{fig:crit_curve3}. The values of $J_{\rm crit}$ reproduced from the previous one-zone simulations by \citet{wolcott-green17} and \citet{agarwal16} are also shown in the Table as a comparison. 

We have obtained $J_{\rm crit}$ from our three simulated haloes. The value of $J_{\rm crit}$ varies from halo to halo, but the variations are within a factor of two. This is shown as a range of values in Table\,3. By comparing $J_{\rm crit}$ calculated in 3D simulations applying the $\lambda_{\rm Jeans}$ approximation, our $J_{\rm crit}$ with T5 is consistent with that from \citet{shang10}. It is about 30\% higher in comparison with \citet{hartwig15}. Our $J_{\rm crit}$ with T4 is consistent with that from \citet{shang10}, but is smaller by one order of magnitude with that from \citet{latif14b}. 

Using different $\lambda$ approximation, the values of $J_{\rm crit}$ differ by up to two orders of magnitude. In our calculations, the $J_{\rm crit}$ in $\lambda_{\rm Sob}$ approximation is smaller than that in $\lambda_{\rm Jeans}$ by almost two orders of magnitude. In addition, for a softer spectrum, e.g., T4, the values of $J_{\rm crit}$ are smaller than the values derived from a harder spectrum with T5, again by up to two orders of magnitude. For the softer spectrum, the radiation field above $0.76$\,eV lead to an increase in $k_{\rm H^-}$. With increasing $k_{\rm H^-}$, the required $k_{\rm H_2}$ or $J_{\rm crit}$ to suppress the $\rm H_2$ cooling decreases. 

Intensities estimated from our 3D simulations tend to be larger compared to those obtained in one-zone simulations. For example, the values of $J_{\rm crit, T5}$ from one-zone simulations by \citet{sugimura14} and \citet{agarwal16} are one order of magnitude smaller than those calculated from 3D simulations by \citet{shang10} and by our work. The difference in $J_{\rm crit}$ between the one-zone and 3D simulations has been also found by \citet{hartwig15} and \citet{latif14b}, who used the single-temperature blackbody spectral shape.

Previous studies which obtained $J_{\rm crit}$, used the single-temperature blackbody spectra. Consequently, these studies obtained only a single point each on the $k_{\rm H_2} - k_{\rm H^-}$ plane. By comparing $J_{\rm crit}$ obtained from previous studies with our values from the critical curve, we find the critical curve provides a more general and compelling way to determine the critical intensity. The single blackbody results from previous works have been reproduced using this curve.

%%%%%%%%%%%%%%%%%%%%%%%%%%%%

\section{Discussion and Concluding Remarks}

We have calculated the conditions for suppressing the formation of H$_2$ in the direct collapse scenario towards the SMBH seeds, within DM haloes.  Using series of 3D numerical simulations we have obtained the critical intensity of the background UV radiation by constructing the critical curves in the $k_{\rm H_2} - k_{\rm H^-}$ parameter plane, separating models with dominant atomic and molecular cooling. We have also shown the dependence of the critical conditions on the choice of the $\rm H_{2}$ column density approximation in the self-shielding calculation. Our main findings can be summarized as follows.

\begin{itemize}
	
\item  We have verified that there exists a critical curve in the $k_{\rm H_2} - k_{\rm H^-}$ parameter plane, above which the $\rm H_{2}$ cooling is suppressed, and the atomic cooling dominates. 

\item We have provided a fitting formula for this critical curve and found that the fitted curve based on the 3D numerical simulations differs substantially from that obtained in the one-zone simulations, both in its position in the $k_{\rm H_2} - k_{\rm H^-}$ parameter plane and in its shape.

\item We have compared the critical curves calculated in 3D simulations using three different $\rm H_{2}$ column density approximations in the self-shielding calculation, $\lambda_{\rm Jeans}$, $\lambda_{\rm Jeans25}$ and $\lambda_{\rm Sob}$. These approximations correspond to the Jeans length, a fraction of the Jeans length and the Sobolev length, respectively. We find that the characteristic lengthscale for shielding can be improved by using $\lambda_{\rm Jeans25}$, which is four times smaller than the local Jeans length.  

\end{itemize}

The direct collapse models involve the gas accretion within the DM haloes, resulting in the SMBH seeds of $\sim 10^4 - 10^6\,\rm M_\odot$. It circumvents the difficulties associated with the growth of the Pop\,III black hole remnants from stellar masses to the SMBH masses found in the galactic centers.  

To sustain a high inflow rate of $\sim 0.1-1\,\rm M_{\odot}\,yr^{-1}$, the $\rm H_{2}$ formation should be inhibited in the primordial gas. To prevent accretion flow fragmentation induced by the $\rm H_{2}$ cooling, the halo must be exposed to the background UV radiation whose intensity exceeds $J_{\rm crit}$. To obtain the $J_{\rm crit}$, typically, a single blackbody or power-law spectra have been assumed in the literature to model the background radiation. However in realistic situations, this radiation field is time-dependent, and a simple spectral shape model cannot capture all the intricacies associated with the flux variability, anisotropy and changing spectral shape. Therefore, it is advantageous to use an alternative approach and deal with the critical intensity that is defined in a more general way, by a combination of $k_{\rm H_2}$ and $k_{\rm H^-}$. There is no need to make any initial assumptions on the properties of the underlying radiation.

We have tested this approach in the fully 3D simulations, and found that a critical curve can exist in the $k_{\rm H_2} - k_{\rm H^-}$ parameter plane. The critical curve position and shape is strongly affected by replacing the one-zone with more realistic simulations in the 3D. Moreover, we have successfully applied a new fitting formula to the critical curve in this parameter space, and compared this curve to those obtained in the one-zone simulations. The main outcome of this comparison is that the critical curve in the 3D simulations lies substantially higher than that from the one-zone simulations, and the required LW flux is higher by up to two orders of magnitude for the same rate $k_{\rm H^-}$.

Our analysis also includes the $\rm H_{2}$ column density approximations for the self-shielding factor calculation. As the gas flows inwards, its density increases, and so is the $\rm H_{2}$ number density. When the $\rm H_{2}$ column densities increase (e.g., $N_{\rm H_2} > 10^{14}\,\rm cm^{-2}$), the photo-dissociation is suppressed because the region becomes optically-thick for the Lyman-Werner photons. The treatment of the gas cooling, therefore, depends on the $\rm H_{2}$ self-shielding approximation. The three cases considered here, $\lambda_{\rm Jeans}$, $\lambda_{\rm Jeans25}$, and $\lambda_{\rm Sob}$, which approximate the characteristic length, provide the column densities, some of which differ from the actual column densities estimated directly from the simulations. The $\lambda_{\rm Jeans}$ approximation overestimates the shielding, while the $\lambda_{\rm Sob}$ approximation significantly underestimates it. We find that $\lambda_{\rm Jeans25}$ suggested by \citet{wolcott-green17} yields the most accurate approach to the true characteristic self-shielding. 

In summary, the 3D simulations in tandem with the $\lambda_{\rm Jeans25}$ approximation for the column density, provide a substantial improvement over the one-zone simulation with fixed spectral shapes of the background UV radiation.

\section*{Acknowledgements}
We thank the Enzo and YT support team for help. All the analysis has been conducted using yt \citep{turk11}, http://yt-project.org/.
Y.L. acknowledges the support from NSFC grant No. 11903026. 
This work has been partially supported by the Hubble Theory grant HST-AR-14584 (to I.S.), and by JSPS KAKENHI grant 16H02163 (to I.S.) and 17H01111 (to K.N.). 
I.S. and K.N. are grateful for a generous support from the International Joint Research Promotion Program at Osaka University. The STScI is operated by the AURA, Inc., under NASA contract NAS5-26555. T.F. acknowledges the support from the National Key R\&D Program of China No. 2017YFA0402600, and NSFC grants No. 11525312, 11890692. Numerical simulations have been performed on Tianhe-2 at the National Supercomputer Center in Guangzhou, on Supercomputer at the Shanghai Astronomical Observatory, as well as on the LCC Linux Cluster of the University of Kentucky.

%%%%%%%%%%%%%%%%%%%%%%%%%%%%%%%%%%%%%%%%%%%%%%%%%%



\bibliographystyle{mnras}
\bibliography{ms} % if your bibtex file is called example.bib

%%%%%%%%%%%%%%%%%%%%%%%%%%%%%%%%%%%%%%%%%%%%%%%%%%

%%%%%%%%%%%%%%%%% APPENDICES %%%%%%%%%%%%%%%%%%%%%

%\appendix



%%%%%%%%%%%%%%%%%%%%%%%%%%%%%%%%%%%%%%%%%%%%%%%%%%

% Don't change these lines
\bsp	% typesetting comment
\label{lastpage}
\end{document}